\documentclass[12pt]{iopart}
\usepackage{graphicx}

\let\Otemize =\itemize
\let\Onumerate =\enumerate
\let\Oescription =\description
\def\Nospacing{\itemsep=0pt\topsep=0pt\partopsep=0pt\parskip=0pt\parsep=0pt}
\def\Topspac{\vspace{-0.4\baselineskip}}
\def\Botspac{\vspace{-0.2\baselineskip}}
\newenvironment{Itemize}{\Topspac\Otemize\Nospacing}{\endlist\Botspac}

\newcommand{\sqrtsNN}{\sqrt{s_{\scriptscriptstyle \rm NN}}}
\newcommand{\av}[1]{\left\langle #1 \right\rangle}

\newcommand{\gev}{\mathrm{GeV}}
\newcommand{\tev}{\mathrm{TeV}}

\newcommand{\mum}{\mathrm{\mu m}}

\renewcommand{\d}{\mathrm{d}}

\renewcommand{\pt}{p_{\rm t}}

\newcommand{\dEdx}{{\rm d}E/{\rm d}x}
\newcommand{\RAA}{R_{\rm AA}}

\begin{document}

\title[Heavy-flavour production in Pb--Pb collisions at the LHC with ALICE]{Heavy-flavour production in Pb--Pb collisions at the LHC, measured with the ALICE detector}

\author{Andrea Dainese, for the ALICE Collaboration}

\address{INFN -- Sezione di Padova, Padova, Italy}
\ead{andrea.dainese@pd.infn.it}

\begin{abstract}
We present the first results from the ALICE experiment 
on the nuclear modification factors for heavy-flavour
hadron production in Pb--Pb collisions at $\sqrt{s_{\rm NN}}=2.76$~TeV.
Using proton--proton and lead--lead collision samples at $\sqrt{s}=7$~TeV and
$\sqrt{s_{\rm NN}}=2.76$~TeV, respectively, 
nuclear modification factors $R_{\rm AA}(p_{\rm t})$ were measured for D mesons
at central rapidity (via displaced decay vertex reconstruction), and for electrons and muons, 
at central and forward rapidity, respectively.
\end{abstract}


\section{Introduction}

The ALICE experiment~\cite{aliceJINST} studies nucleus--nucleus and 
proton--proton collisions at the Large Hadron Collider (LHC), with the
main goal of investigating the properties of the high-density, colour-deconfined,
state of strongly-interacting matter that is expected to be formed
in Pb--Pb collisions. 
The first Pb--Pb data were collected in November 2010 at 
a centre-of-mass energy $\sqrtsNN=2.76~\tev$ per nucleon--nucleon collision.

Heavy-flavour particles, abundantly produced at LHC energies,
are regarded as effective probes of the conditions of the system (medium) formed 
in nucleus--nucleus collisions:
\begin{Itemize}
\item open charm and beauty hadrons should be sensitive to the energy density,
through the mechanism of in-medium energy loss of heavy quarks;
\item quarkonium states should be sensitive to the initial temperature of the
system, through their dissociation due to colour screening.
\end{Itemize}
The first ALICE results on charmonium production in Pb--Pb collisions are described
in~\cite{ginesphilippe}. In this report we present first measurements of the medium-induced
suppression of high transverse momentum ($\pt$) heavy-flavour hadrons.

The nuclear modification factor $R_{\rm AA}$ of particle $\pt$ distributions
is well-established as a sensitive observable 
for the study of the interaction of hard partons 
with the medium. This factor is defined as 
the ratio of the $\pt$ spectrum measured in nucleus--nucleus (AA)
to that expected on the basis of the proton--proton spectrum scaled 
by the number $N_{\rm coll}$ of binary 
nucleon--nucleon collisions in the nucleus--nucleus collision: 
\begin{equation}
R_{\rm AA}(\pt)=
{1\over \av{N_{\rm coll}}} \cdot 
{\d N_{\rm AA}/\d\pt \over 
\d N_{\rm pp}/\d\pt} = 
{1\over \av{T_{\rm AA}}} \cdot 
{\d N_{\rm AA}/\d\pt \over 
\d\sigma_{\rm pp}/\d\pt}\,,
\end{equation}
where the AA spectrum corresponds to a given collision-centrality class and 
$\av{T_{\rm AA}}$ is the average nuclear overlap function resulting for that 
centrality class from the 
Glauber model of the collisions geometry~\cite{glauber}.
A strong suppression in the $\RAA$ of charged particles was observed in 
Pb--Pb collisions at the LHC~\cite{raapaper}.
Due to the 
QCD nature of parton energy loss, quarks are predicted to lose less
energy than gluons (that have a higher colour charge) and, in addition, 
the `dead-cone effect' and other mechanisms are expected to reduce the energy loss of massive 
quarks with respect to light partons (light-flavour quarks and gluons)~\cite{dk,asw,whdg}. 
Therefore, one should observe a pattern
of gradually decreasing $\RAA$ suppression when going from the mostly 
gluon-originated
light-flavour hadrons (e.g. pions) 
to D and to B mesons~\cite{adsw}: 
$\RAA^\pi<\RAA^{\rm D}<\RAA^{\rm B}$. 
The measurement and comparison of these different medium probes 
provide a unique test of the 
colour-charge and mass dependence of parton energy loss.

The ALICE experiment was designed to detect heavy-flavour hadrons
over a wide phase-space coverage and in various decay channels, as we will 
discuss in section~\ref{sec:hf}.
The production of heavy-flavour probes was measured in proton--proton collisions at 
$\sqrt{s}=7~\tev$ and compared to perturbative QCD (pQCD) predictions. These results,
reported in section~\ref{sec:pp}, provide a well-defined, calibrated, reference for 
$\RAA$ measurements. The reference at the Pb--Pb $\sqrtsNN$ energy is obtained by
a pQCD-driven scaling (section~\ref{sec:scaling}). The Pb--Pb analysis results and 
the nuclear modification factors of $\rm D^0$ and $\rm D^+$ mesons, and of 
electrons and muons from heavy-flavour decays are presented in 
section~\ref{sec:pbpb}. More details on these analyses can be found 
in~\cite{andrea,silvia,yvonne,xiaoming}.

\section{Heavy-flavour production measurements in ALICE}
\label{sec:hf}

The heavy-flavour detection capability of the ALICE detector,
described in detail in~\cite{aliceJINST}, is mainly provided by:
\begin{Itemize}
\item Tracking system; the silicon Inner Tracking System (ITS) and 
the Time Projection Chamber (TPC)
embedded in a magnetic field of $0.5$~T, allow track reconstruction in 
the pseudo-rapidity range $|\eta|<0.9$ 
with a momentum resolution better than
4\% for $\pt<20~\gev/c$ 
and a transverse impact parameter\footnote{The transverse impact parameter,
$d_0$, is defined as the distance of closest approach of the track to the 
interaction vertex, in the plane transverse to the beam direction.} 
resolution better than 
$75~\mum$ for $\pt>1~\gev/c$ with a high-$\pt$ saturation value of $20~\mum$.
\item Particle identification system; charged hadrons are separated via 
their specific energy deposit $\dEdx$ in the TPC and via time-of-flight measurement in the 
Time Of Flight (TOF) detector; 
electrons are identified at low $\pt$ ($<6~\gev/c$) via TPC $\dEdx$ and 
time-of-flight, and at intermediate and high $\pt$ ($>2~\gev/c$) in the dedicated Transition Radiation Detector (TRD) and
in the Electromagnetic Calorimeter (EMCAL); 
muons are identified in the muon 
spectrometer covering the pseudo-rapidity range $-4<\eta<-2.5$\footnote{
Since the colliding particles, pp and Pb--Pb, are symmetric in mass, we indicate
the muon spectrometer acceptance using positive values in the following.}. 
\end{Itemize}
The main detection modes are listed below.
\begin{Itemize}
\item Charm hadronic decays in $|y|<0.5$ using displaced vertex
identification: 
$\rm D^0 \to K^-\pi^+$,  
$\rm D^+ \to K^-\pi^+\pi^+$, and $\rm D^{*+}\to D^0\pi^+$. Other channels
and hadrons are under study: $\rm D^0 \to K^-\pi^+\pi^-\pi^+$,
$\rm D_s^+ \to K^-K^+\pi^+$, and $\rm \Lambda_c^+ \to p K^-\pi^+$.
\item Leptons from heavy-flavour decays: 
inclusive ${\rm D/B\to e}+X$ in $|y_{\rm e}|<0.8$ and ${\rm D/B\to\mu}+X$ in $2.5<y_{\mu}<4$; 
${\rm B\to e}+X$ using displaced electron identification (only in pp for the moment).
${\rm B\to J/\psi\,(\to e^+e^-)}+X$ is also under study.
\end{Itemize} 

The results that we present are obtained from data recorded with minimum-bias trigger
selections in pp and Pb--Pb collisions. These selections were
defined by the presence of signals in either of two scintillator hodoscopes, the V0 detectors, 
positioned in the forward and backward regions of the experiment, or in the silicon pixel barrel
detector, in coincidence with the signals of two beam pick-up counters on both sides
of the interaction region. For muon analyses in pp collisions, a second selection
requests, in addition to the minimum-bias condition, 
a muon trigger signal ($\pt>0.5~\gev/c$). The results presented correspond to 
about 100--180 million (depending on the analysis) pp minimum-bias triggers and 
about 17 million Pb--Pb minimum-bias triggers. For pp collisions, the heavy flavour production 
cross sections are normalized relative to the minimum-bias trigger cross section, which is 
derived using a van-der-Meer scan technique.
Pb--Pb collision-centrality classes
are defined in terms of percentiles of the distribution of 
the sum of amplitudes in the V0 scintillator detectors.

\section{Calibrating heavy-flavour probes with pp collisions at $\sqrt{s}=7~\rm TeV$}
\label{sec:pp}

\begin{figure}[!t]
  \begin{center}
  \includegraphics[width=0.32\textwidth]{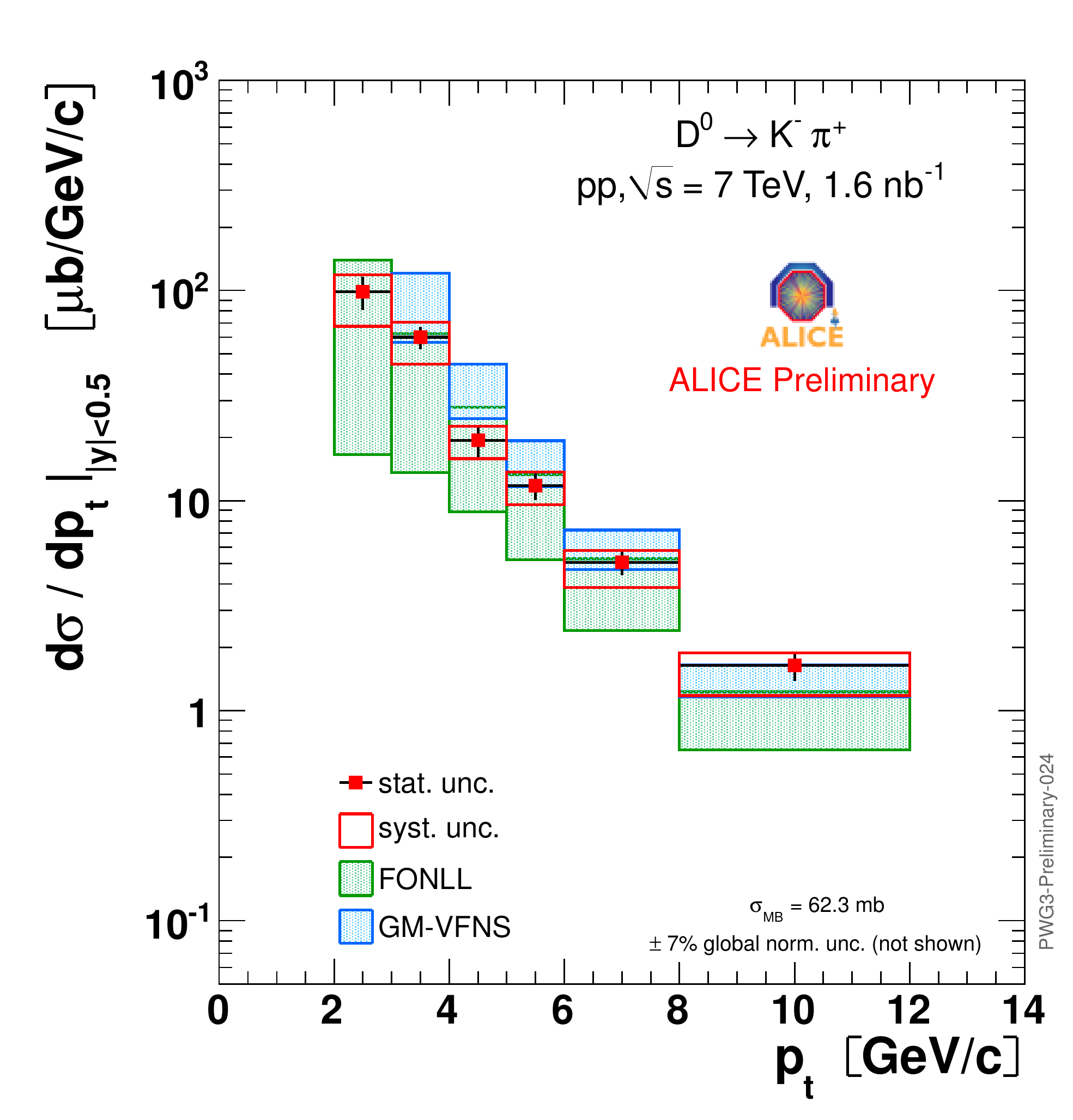}
  \includegraphics[width=0.32\textwidth]{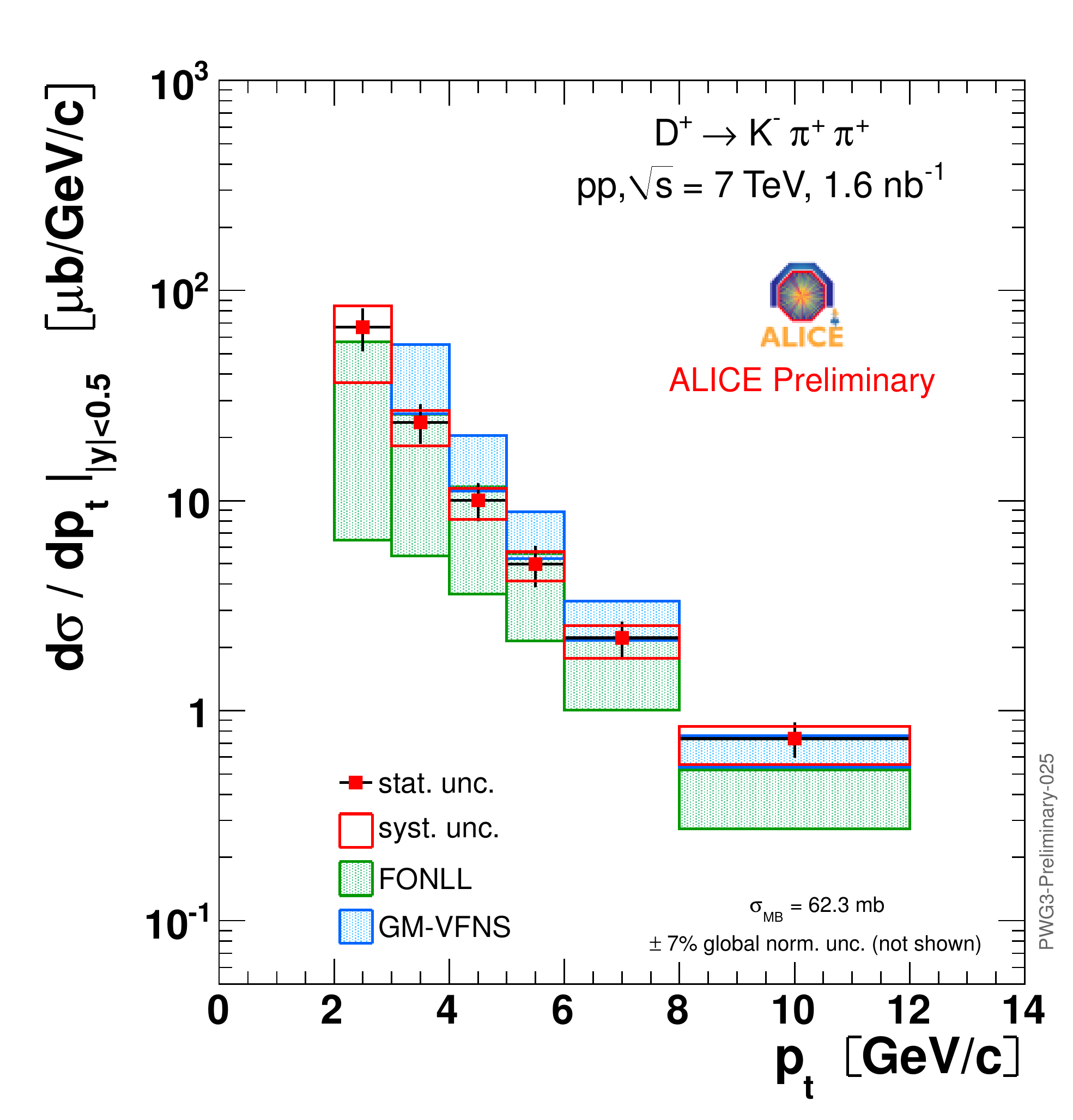}
  \includegraphics[width=0.32\textwidth]{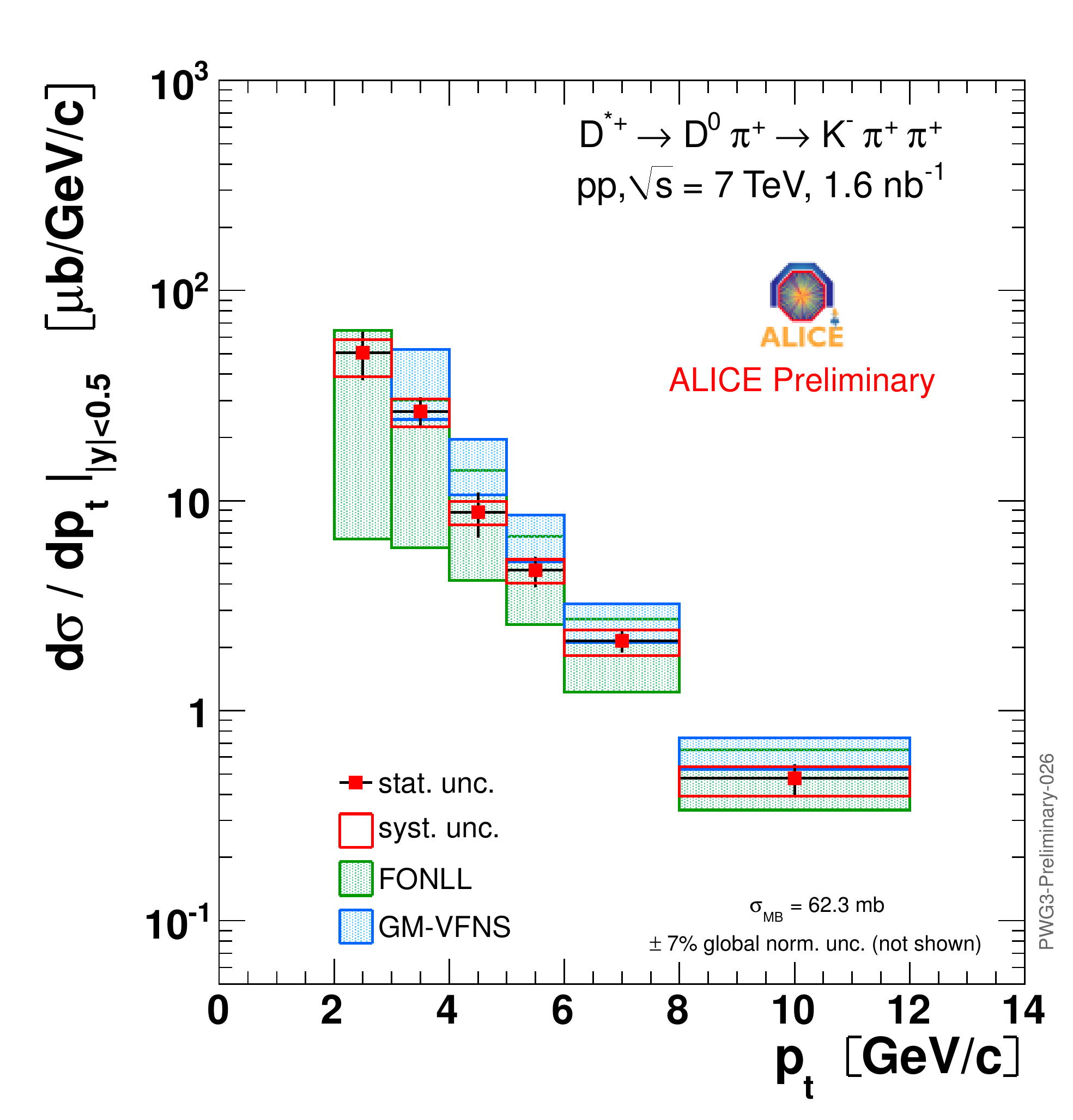}
  \caption{$\rm D^0$, $\rm D^+$, and $\rm D^{*+}$ $\pt$-differential production 
cross sections in $|y|<0.5$ in pp collisions at $\sqrt{s}=7~\tev$, compared to 
pQCD calculations~\cite{fonllpriv,gmvfnspriv}.}
\label{fig:Dpp}
\end{center}
\end{figure}

The detection strategy for D mesons at central rapidity
is based, for both pp and Pb--Pb collisions,  
on the selection of displaced-vertex topologies, i.e. separation of tracks
from the secondary vertex from those from the primary vertex, 
large decay length (normalized to its estimated uncertainty),
and good alignment between the reconstructed D meson momentum 
and flight-line. 
The identification of the charged kaon in the TPC and TOF detectors helps to further reduce the
background at low $\pt$. 
An invariant-mass analysis is then used to extract the raw signal 
yield, to be then corrected for detector acceptance and 
for PID, selection and reconstruction efficiency, evaluated from a detailed detector simulation. 
The contamination of D mesons from B meson
decays is estimated to be of about 15\%, using the beauty production cross
section predicted by the FONLL (fixed-order next-to-leading log) 
calculation~\cite{fonllpriv} and the detector simulation, 
and it is subtracted from the measured raw $\pt$ spectrum, before applying the efficiency 
corrections.
The $\rm D^0$, $\rm D^+$, and $\rm D^{*+}$ $\pt$-differential production 
cross sections in $|y|<0.5$ are shown in Fig.~\ref{fig:Dpp}. Theoretical predictions based on pQCD calculations (FONLL~\cite{fonllpriv} and GM-VFNS~\cite{gmvfnspriv}) are in agreement 
with the data.

\begin{figure}[!t]
  \begin{center}
\includegraphics[width=0.38\textwidth]{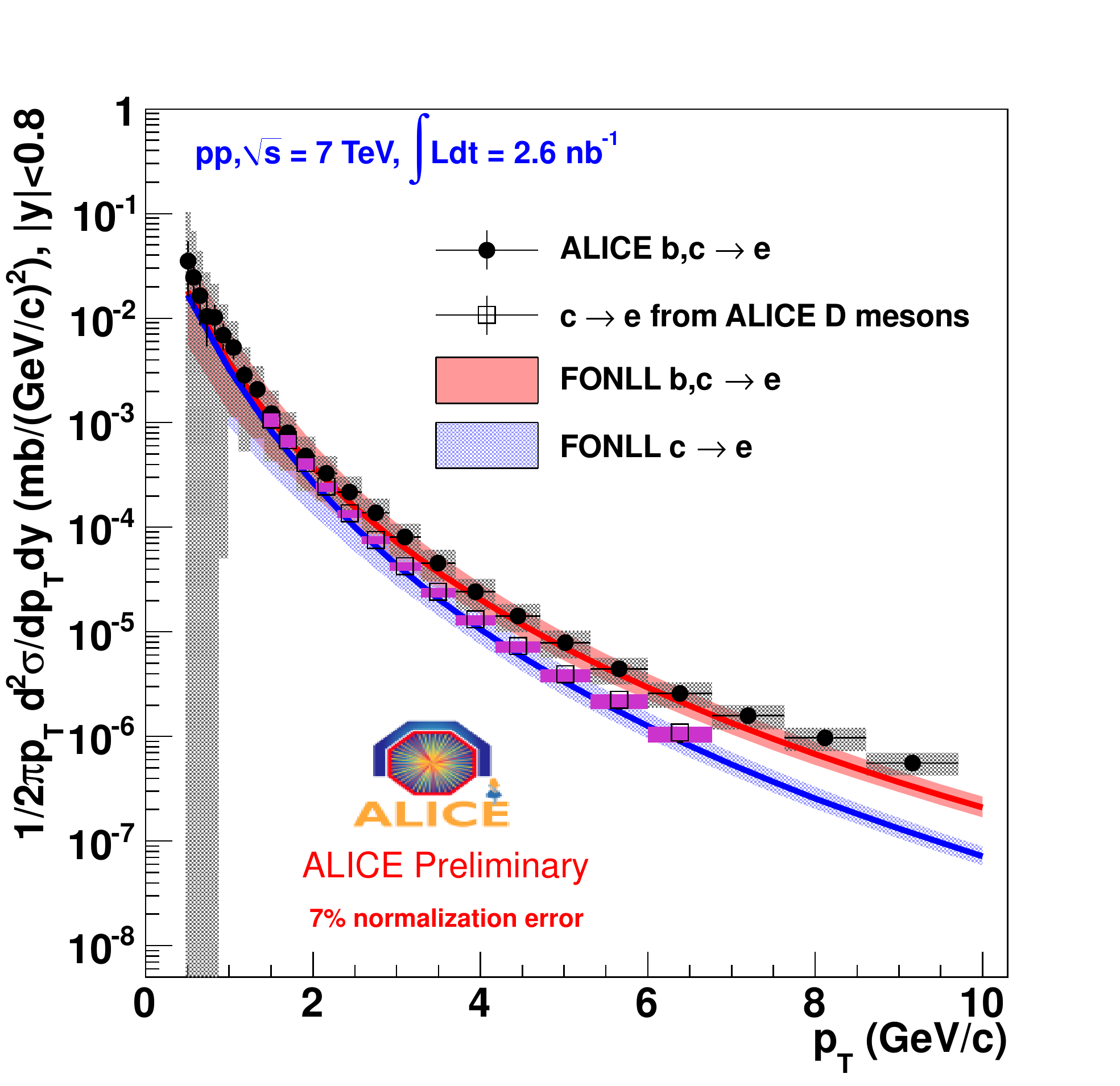}
  \includegraphics[width=0.32\textwidth]{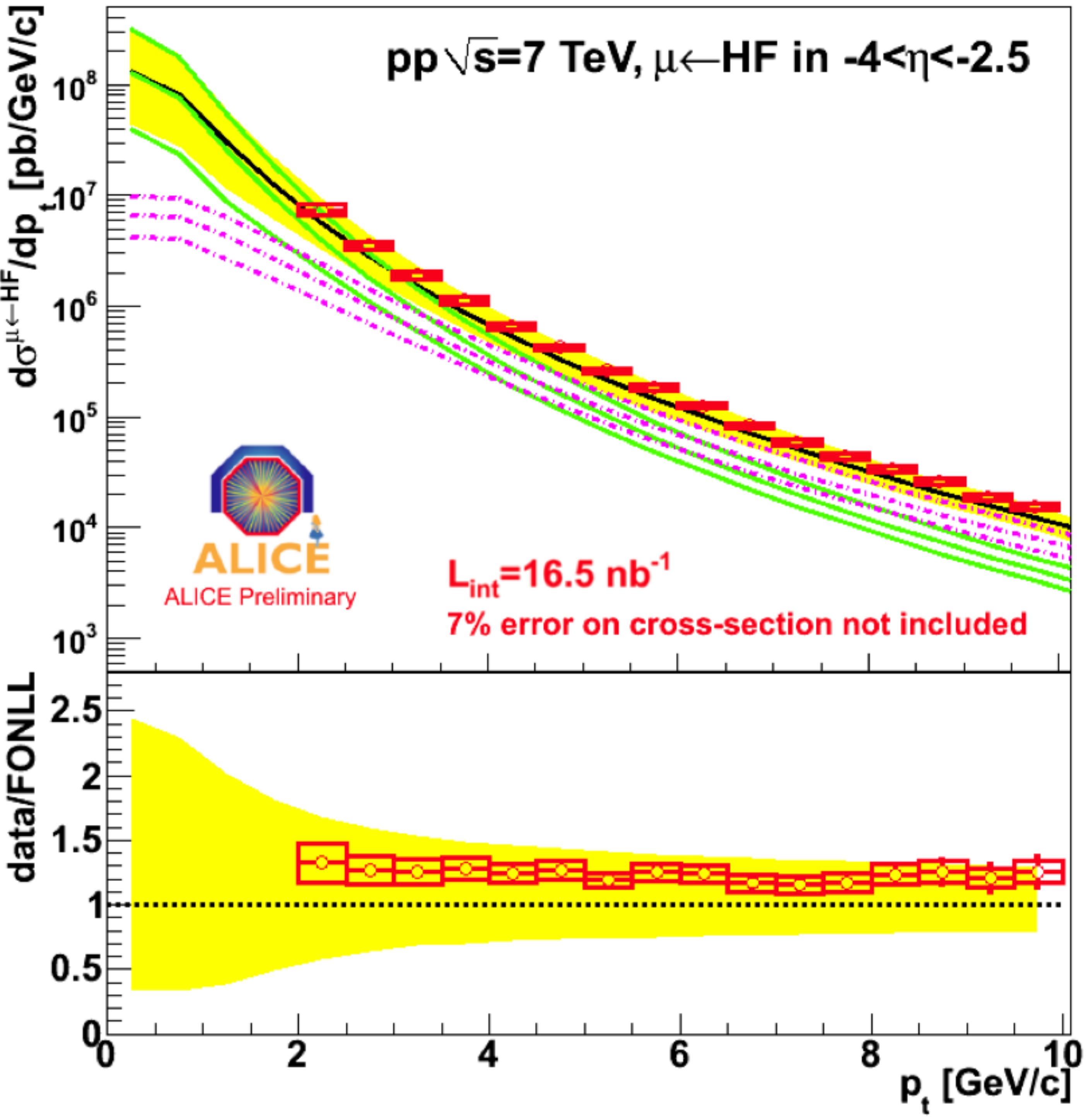}
  \caption{$\pt$-differential production 
cross sections of heavy-flavour decay electrons in $|y|<0.8$ (left) and muons in $2.5<y<4$
(right) in pp collisions at $\sqrt{s}=7~\tev$, compared to 
pQCD calculations~\cite{fonllpriv}. The left panel shows also the cross section of electrons obtained by applying the decay kinematics to the D cross sections in Fig.~\ref{fig:Dpp}.}
\label{fig:leptpp}
\end{center}
\end{figure}

At central rapidity, heavy flavour production is measured also using semi-electronic decays.
The basis of this measurement is a robust electron identification. 
For pp collisions, this uses the signals in the TPC, TOF, and TRD detectors,
as detailed in~\cite{silvia} (the EMCAL information will also be exploited in the near
future).
The residual pion contamination in the electron sample is measured, 
and then subtracted, by fitting with 
a two component function the TPC $\dEdx$ distribution in narrow momentum slices.
The $\pt$-differential cross section of electrons from 
charm and beauty particle decays is obtained by subtracting from the 
efficiency-corrected inclusive spectrum a ``cocktail" of background electrons. The components of the cocktail are 
electrons from light-flavour hadron decays (mainly
    $\pi^0$ Dalitz decays, in addition to $\eta$, $\rho$, $\omega$, and $\phi$ decays), 
    and $\gamma$
    conversions in the beam pipe and innermost pixel layer. These inputs
 are determined from the measured pion cross section, 
 using simulations and $m_{\rm t}$ scaling~\cite{silvia}.
Figure~\ref{fig:leptpp} (left) shows the resulting heavy-flavour decay electrons cross section in 
$|y|<0.8$, compared to the corresponding FONLL prediction~\cite{fonllpriv}
and to the cross section of charm decay electrons, obtained by applying the decay 
kinematics to the D cross sections.

Heavy flavour production at forward rapidity is measured using the single-muon $\pt$ 
distribution.
The extraction of the heavy-flavour contribution from the single muon spectra requires the subtraction of three main sources of background: muons from the decay-in-flight of light hadrons (decay muons); muons from the decay of hadrons produced in the interaction with the front absorber (secondary muons); hadrons that punch through the front absorber.
The last contribution, about 20\% for $\pt>2~\gev/c$, can be efficiently rejected by requiring the matching of the reconstructed tracks with the tracks in the trigger system. Due to the lower mass of the parent particles, the
light-hadron decay muons have a softer transverse momentum than the heavy-flavour muons, 
and dominate the low-$\pt$ region. In the pp analysis this background is subtracted using
simulations, as detailed in~\cite{xiaoming}. 
The charm and beauty decay muon cross section $\d\sigma/\d\pt$ in the range 
 $2.5<y<4$ is presented in Fig.~\ref{fig:leptpp} (right). The corresponding 
FONLL~\cite{fonllpriv} prediction agrees with our data
and indicates that beauty decays are the dominant contribution for $\pt>6~\gev/c$.

\section{pp references at $\sqrt{s}=2.76~\rm TeV$ via pQCD-driven $\sqrt{s}$-scaling}
\label{sec:scaling}

The reference pp cross sections used for the determination of the nuclear 
modification factors were obtained by applying a $\sqrt{s}$-scaling~\cite{scaling} 
to the cross sections 
measured at $\sqrt{s}=7~\tev$. 
The scaling factors for each of the three observables (D mesons, muons, electrons)
were defined as the ratios of the cross sections from 
the FONLL pQCD calculation at $2.76$ and $7~\tev$. The assumption that was
made is that the following quantities do not vary with $\sqrt{s}$: 
a) the pQCD scale values (factorization and renormalization scales)
and b) the c and b quark masses used in the calculation. The theoretical uncertainty on the
scaling factor was evaluated by considering the envelope of the scaling factors resulting from 
different values of the scales and heavy quark masses. This uncertainty is similar for 
D mesons and for heavy-flavour decay leptons at the two rapidities and it ranges from
25\% at $\pt=2~\gev/c$ to less that 10\% above $10~\gev/c$~\cite{scaling}.

For D mesons, the procedure was validated by scaling the ALICE data 
to Tevatron energy, $\sqrt{s}=1.96~\tev$, and comparing~\cite{scaling} 
to CDF measurements~\cite{cdf}.

The $\rm D^0$ and $\rm D^+$ cross sections were measured, 
though with limited precision of 20--25\% and $\pt<8~\gev/c$, also  
for pp collisions at $\sqrt{s}=2.76~\tev$ collected during a short reference run at the 
same energy as Pb--Pb collisions. These measurements at $\sqrt{s}=2.76~\tev$ 
were found to be in 
agreement with the scaled $7~\tev$ measurements.

\section{Heavy-flavour nuclear modification factors in Pb--Pb at $\sqrt{s_{\scriptscriptstyle\rm NN}}=2.76~\rm TeV$}
\label{sec:pbpb}

\begin{figure}[!t]
  \begin{center}
\includegraphics[width=0.44\textwidth]{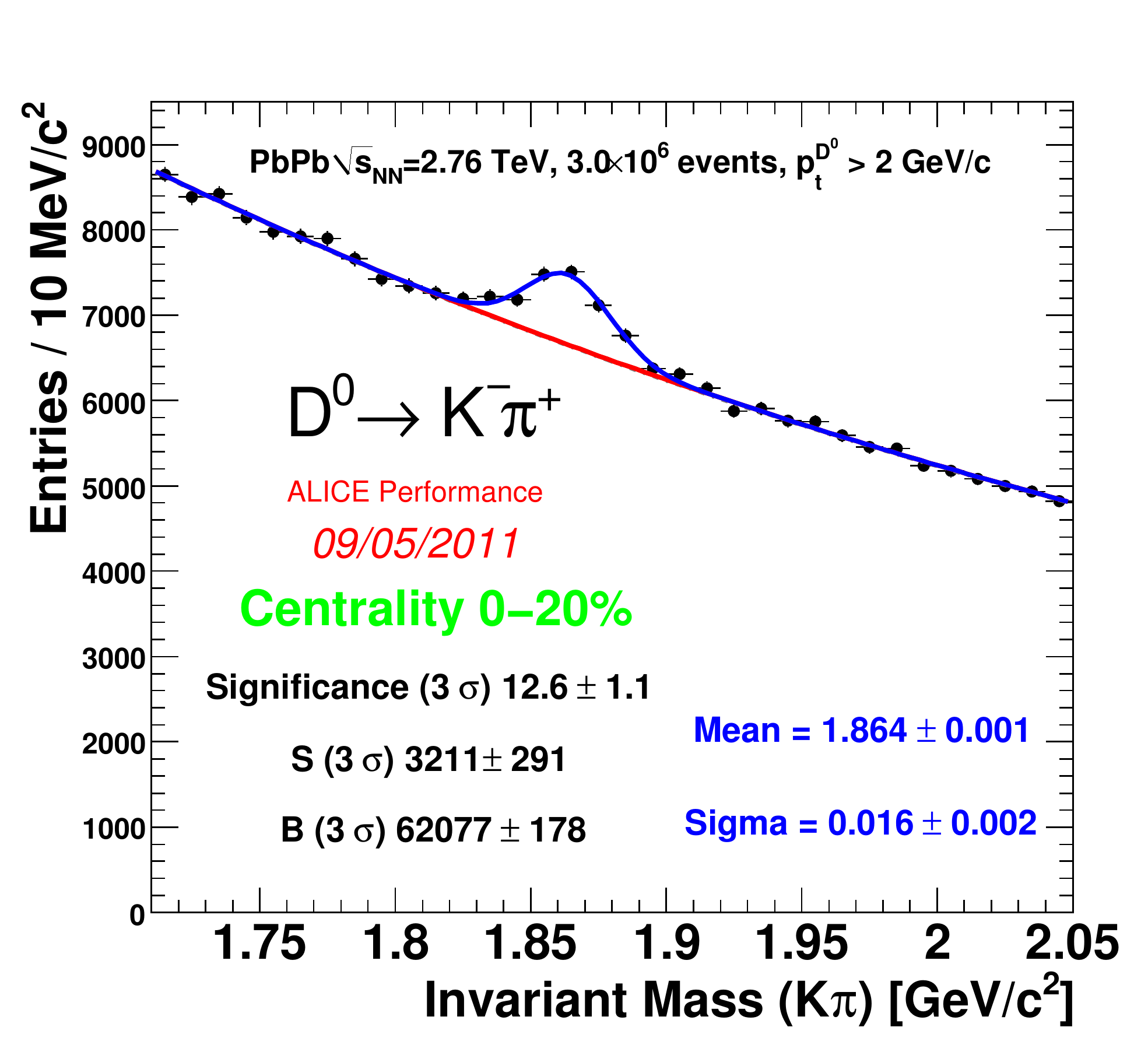}
  \includegraphics[width=0.54\textwidth]{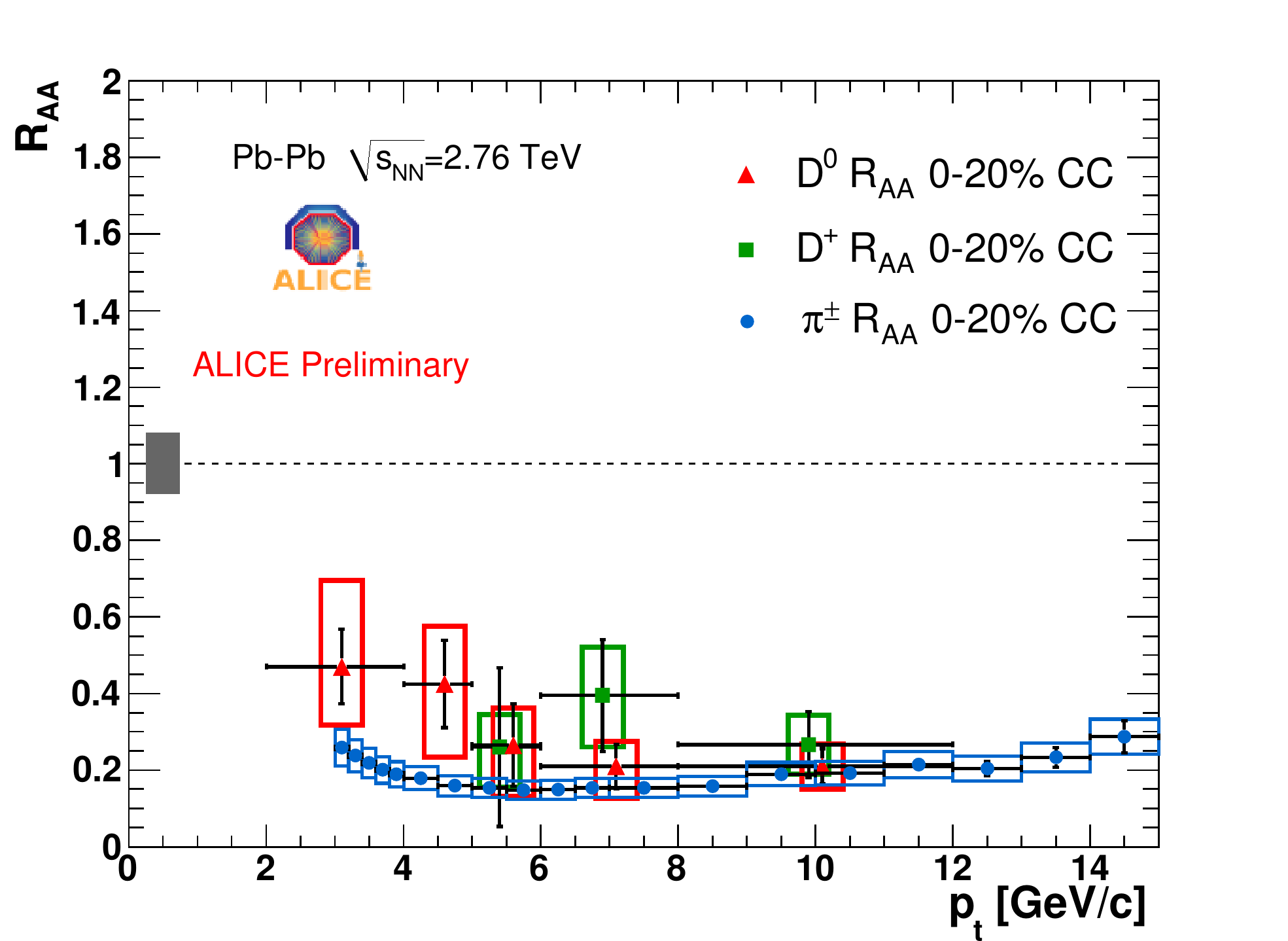}
  \caption{Left: $\rm D^0\to K^-\pi^+$ signal in the $\rm K\pi$ invariant mass distribution
  for central (0--20\%) Pb--Pb. Right: $\RAA$ 
  for $\rm D^0$, $\rm D^+$, and $\pi^+$ in central collisions. Statistical (bars), 
  systematic (empty boxes), and normalization (full box) uncertainties are shown.}
\label{fig:DPbPb}
\end{center}
\end{figure}

D mesons are reconstructed in Pb--Pb collisions~\cite{andrea} 
using the same strategy as for the pp case,
exploiting the vertexing precision and the hadron identification capabilities of the 
ALICE detector. The $\rm D^0\to K^-\pi^+$ signal was measured in five $\pt$ bins
in 2--12~GeV/$c$ and the $\rm D^+\to K^-\pi^+\pi^+$ signal in three bins
in 5--12~GeV/$c$.
An example $\rm K\pi$ invariant mass distribution for 
$\pt>2~\gev/c$ in the 0--20\% centrality class is shown in Fig.~\ref{fig:DPbPb} (left).
The reconstruction and cut selection efficiency, evaluated from detailed detector simulation, 
ranges between 1\ and 10\% for increasing $\pt$ and it does not show a significant dependence
on the collision centrality. The subtraction of feed-down from B meson decays relies on the 
FONLL predictions, as for the pp case. This contribution is of about 10--15\%. 
The systematic uncertainty related to the FONLL theoretical uncertainty is correlated in the 
pp and Pb--Pb spectra, thus it partly cancels in the $\RAA$ ratio. An additional source of 
systematic uncertainty arises from the unknown nuclear modification of
B meson production. To estimate this uncertainty, the prompt D meson $\RAA$
was evaluated for the range of hypotheses $0.3<\RAA^{\rm B}/\RAA^{\rm D}<3$ and the 
resulting variation was found to be smaller than 15\%. This is relatively small in comparison 
to the total experimental systematic uncertainty of about 35\%, which is dominated by 
the track reconstruction, particle identification and cut efficiency corrections.
The nuclear modification factor of prompt $\rm D^0$ and $\rm D^+$ mesons
in central 
(0--20\%) Pb--Pb collisions is shown in Fig.~\ref{fig:DPbPb} (right). A strong suppression 
is observed, reaching a factor 4--5 for $\pt>5~\gev/c$ for both D meson species. 
Their $\RAA$ is compatible with 
that of charged pions, also shown, although there seems to be a tendency for
$\RAA^{\rm D}>\RAA^\pi$ at low $\pt$. 

\begin{figure}[!t]
  \begin{center}
\includegraphics[width=0.5\textwidth]{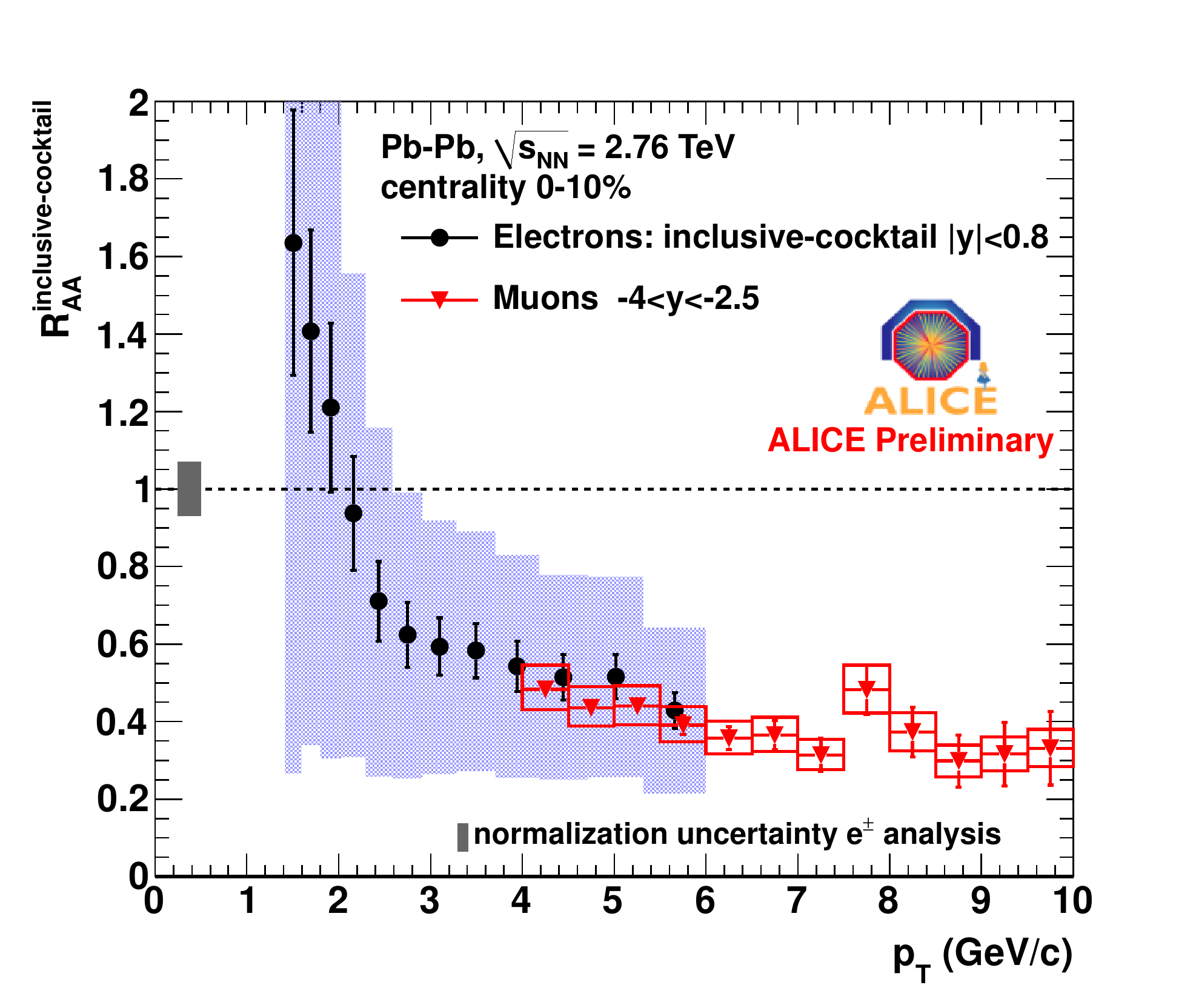}
  \caption{Nuclear modification factors for cocktail-subtracted inclusive electrons 
  at central rapidity and inclusive muons at forward rapidity 
  in central (0--10\%) Pb--Pb collisions. Statistical (bars) and
  systematic (empty boxes) uncertainties are shown.}
\label{fig:leptPbPb}
\end{center}
\end{figure}

\begin{figure}[!t]
  \begin{center}
\includegraphics[width=0.43\textwidth]{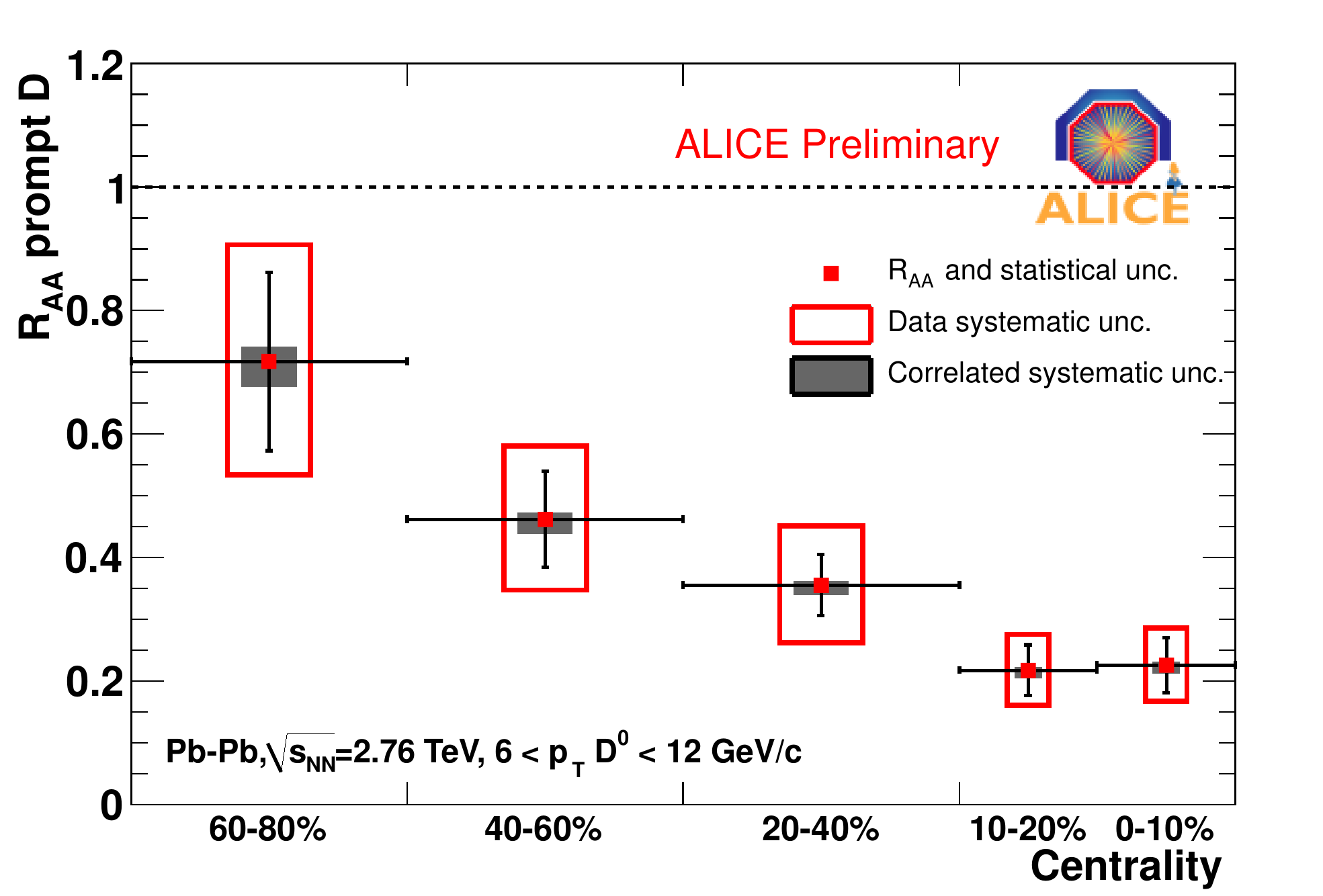}
\\
\vskip-5mm
\includegraphics[width=0.38\textwidth]{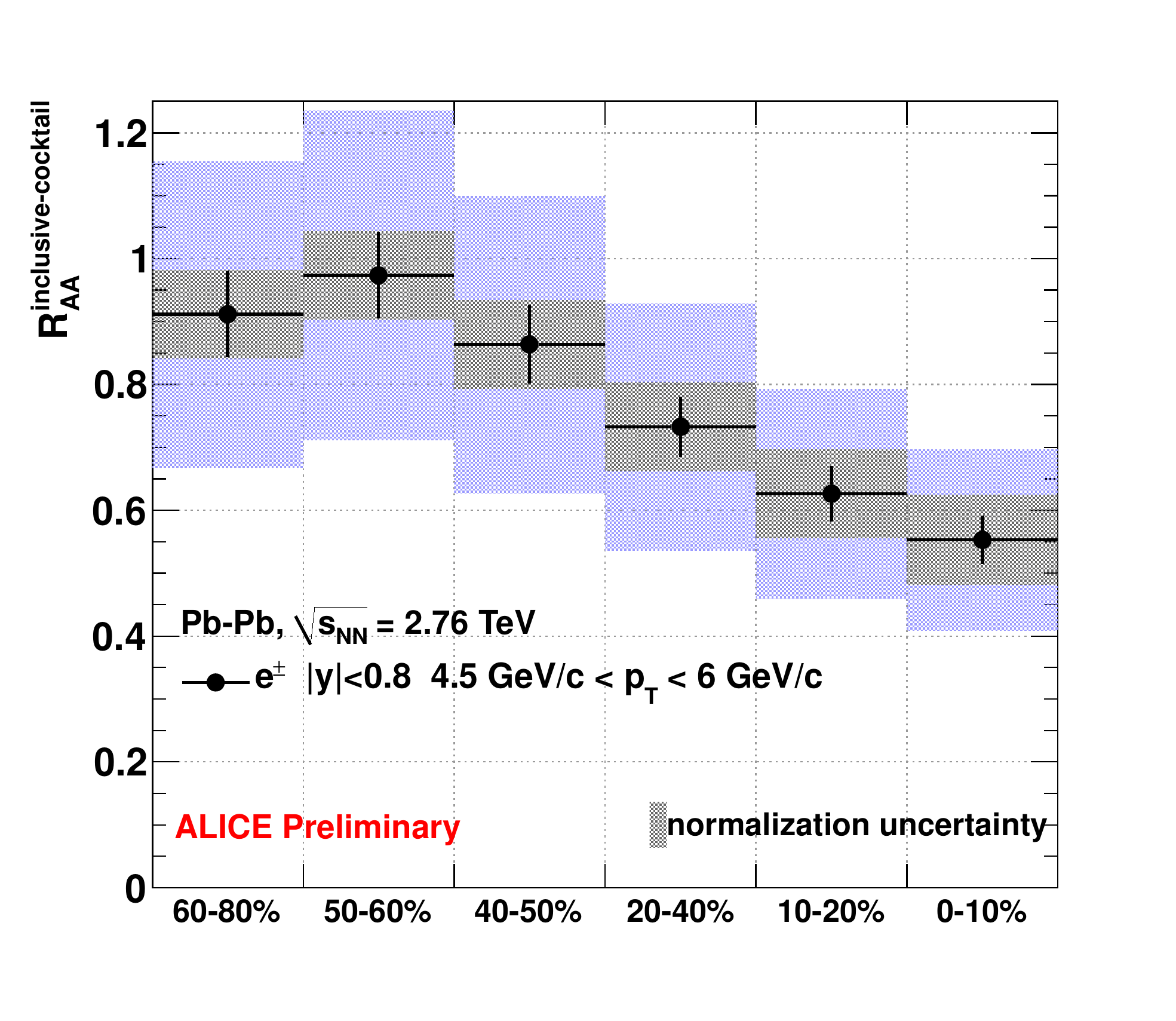}
\includegraphics[width=0.43\textwidth]{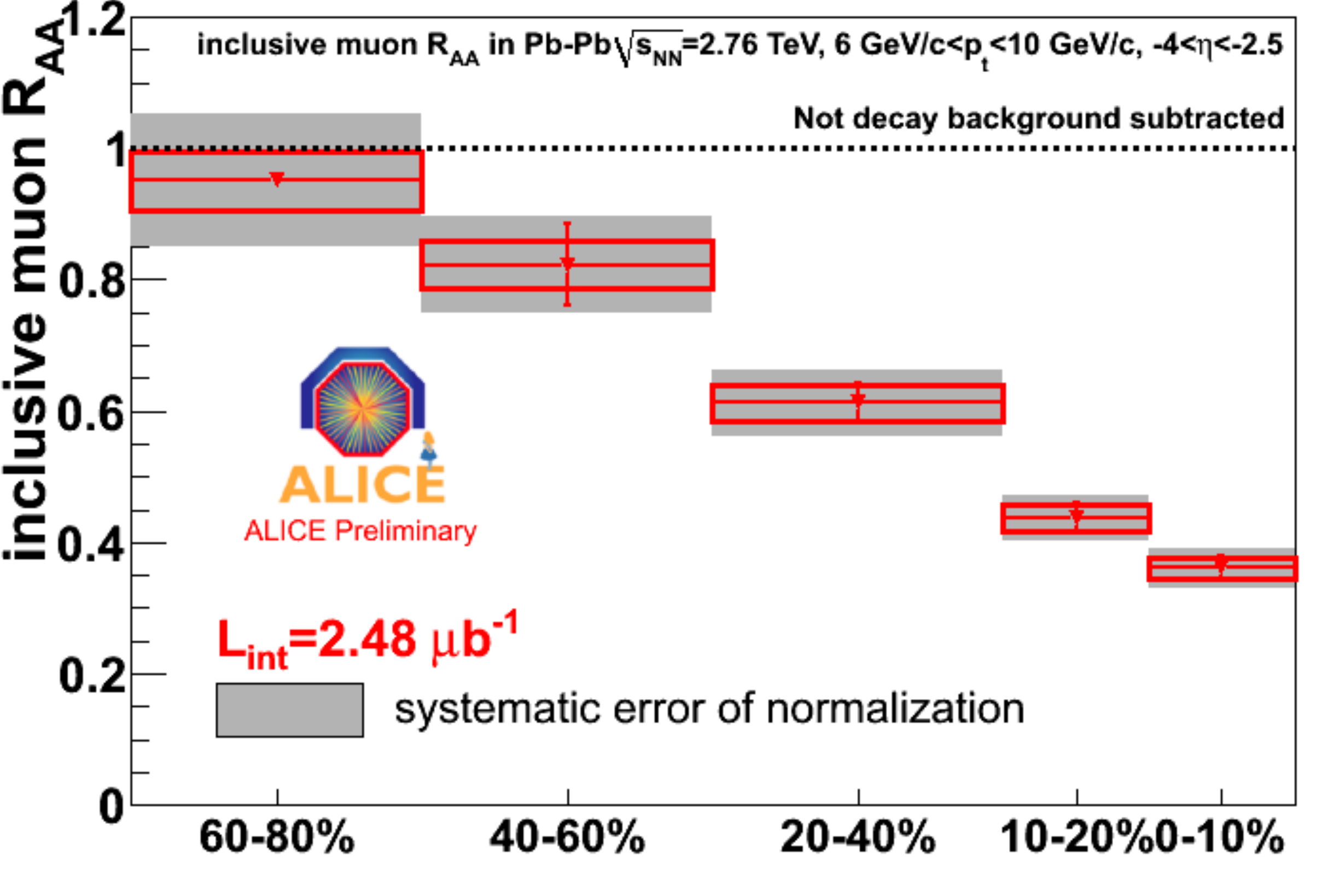}
  \caption{Centrality dependence of $\RAA$
    for $\rm D^0$ mesons (top) 
   for cocktail-subtracted inclusive electrons  
  at central rapidity (bottom-left), and for inclusive muons at forward rapidity (bottom-right).
  The $\pt$ intervals are indicated in the figures.}
\label{fig:centrPbPb}
\end{center}
\end{figure}

Like for D mesons, also for single electrons and single muons the Pb--Pb analysis strategy 
is similar to the pp case. 
Electron identification is based on the TPC and TOF detector signals only, for the moment,
and the electron $\pt$ spectra are limited to the region below 
$6~\gev/c$, to keep the hadron contamination 
below 15\% (as in pp, measured from the TPC ${\rm d}E/{\rm d}x$). The inclusive electron 
spectra, corrected for efficiency and acceptance, and the background cocktail,
based on measured $\pi^\pm$ spectra, are extracted in six centrality classes.
At variance with pp and peripheral Pb--Pb collisions, for semi-central and central Pb--Pb
collisions the comparison with the cocktail shows a hint for an excess of electrons 
in the low-$\pt$ region (up to about $3~\gev/c$),
that increases towards more central collisions~\cite{yvonne}. 
This could be related to 
thermal photons radiated by the hot QCD medium. For $\pt>3$--$4~\gev/c$
the background-subtracted inclusive electron spectrum should be dominated by decays 
of D and B mesons. We use this spectrum to compute $\RAA(\pt)$, 
which is shown for the 0--10\% centrality class in Fig.~\ref{fig:leptPbPb}. A suppression of 
a factor 1.5--4 is observed above $4~\gev/c$. The large systematic uncertainty is dominated
by the electron identification corrections. The same figure shows also the $\RAA$ of 
inclusive muons in $2.5<y<4$. In Pb--Pb the light-flavour decay muon component is not
subtracted as in pp, but we estimated using simulations that this background is 
smaller than 10--15\% for $\pt>6~\gev/c$. The muon $\RAA$ shows a suppression of factor 
about 3 up to $10~\gev/c$, in a momentum range where beauty decay muons should be 
dominant.

Figure~\ref{fig:centrPbPb} shows the centrality dependence of the nuclear modification 
factors for $\rm D^0$ mesons, single electrons, and single muons, 
for $\pt>4$ ($\rm e^+$) and $6$ ($\mu^+$ and $\rm D^0$) $\gev/c$.
For all three cases, the suppression tends to vanish towards peripheral collisions. The 
$\RAA$ of electrons at central rapidity and of muons at forward rapidity is compatible, 
while the $\RAA$ of prompt D mesons is systematically lower. This feature is present in 
the predictions of models in which the energy loss is smaller for b than for c 
quarks~\cite{adsw,zaida}.

\vspace{-0.2cm}

\section{Conclusions}

The first ALICE results on the $\RAA$ 
nuclear modification factors of heavy-flavour
hadrons in Pb--Pb collisions at the LHC imply a strong 
in-medium energy loss for c and b quarks.
The $\rm D^0$ and $\rm D^+$ $\RAA$, 
measured for the first time as a function of $\pt$ and centrality,
is as low as 0.2--0.3 for $\pt>5~\gev/c$ and compatible with the $\RAA$ of pions.
Below $5~\gev/c$, there is an indication for a rise and for $\RAA^{\rm D}>\RAA^\pi$. 
Higher statistics data, expected from the 2011 Pb--Pb run, and comparison data 
in p--Pb collisions should allow to study this region with more precision and disentangle
the initial-state nuclear effects, which could be different for light and heavy flavours.
A large suppression, $\RAA\approx 0.3$--0.4, however systematically larger than for D mesons,
is observed also for inclusive electrons and muons in 
the $\pt$ range above $5~\gev/c$, where beauty decays are dominant according to 
pQCD calculations. The direct measurement of $\RAA$ of electrons from beauty decays
should be possible with the data of the 2011 Pb--Pb run.

\end{document}